%% file: a.tex
\documentclass[11pt,a4paper]{article}

\def\mpt{{{\bf p}\!\!\!/}_T}

\newcommand{\citemsugraandcmssm}{\cite{Chamseddine:1982jx,Barbieri:1982eh,Ibanez:1982ee,Hall:1983iz,Ohta:1982wn,Kane:1993td}}

\input{ourSusydefs}

\usepackage{jheppub}
\usepackage{multirow}
\usepackage{axodraw}
\usepackage{graphics}

\newcommand\squark{{\tilde q}}
\newcommand\gluino{{\tilde g}}

\newcommand\noteOnAreasAndLabels{The area of each
  histogram has been 
  normalised to 1 and labeled `Incl.\ ATLAS' (`Excl.\ ATLAS') if it
  includes (excludes) the ATLAS results.}

\author[a]{B.C.~Allanach,}
\affiliation[a]{Department of Applied Mathematics and Theoretical Physics, Centre for Mathematical Sciences, University of Cambridge, Wilberforce Road, Cambridge CB3 0WA, United Kingdom}
\emailAdd{B.C.Allanach@damtp.cam.ac.uk}

\author[b]{T.J.~Khoo,}
\emailAdd{khoo@hep.phy.cam.ac.uk}

\author[b]{C.G.~Lester,}
\affiliation[b]{Department of Physics,
Cavendish Laboratory,
J J Thomson Avenue,
Cambridge,
CB3 0HE, United Kingdom}
\emailAdd{lester@hep.phy.cam.ac.uk}

\author[b]{S.L.~Williams}
\emailAdd{slw55@cam.ac.uk}

\title{The impact of the ATLAS zero-lepton, jets and missing momentum search on a CMSSM fit}

\abstract{Recent ATLAS data significantly extend the exclusion limits for
  supersymmetric particles. We examine the impact of such data on global fits of the
  constrained minimal supersymmetric standard model (CMSSM) to indirect and
  cosmological data.
  We calculate the likelihood map of the ATLAS search, taking into account 
  systematic errors on the signal and on the background. We validate our
  calculation against the ATLAS determinaton of 95$\%$ confidence level (C.L.)
  exclusion contours. 
  A previous CMSSM global fit is then re-weighted by the 
  likelihood map, which takes a bite at the high 
  probability density region of the global fit, 
  pushing scalar and gaugino masses up. 
}

\newcommand{\twographs}[4]{
\unitlength=1.143in
\begin{picture}(6,2)(0,0)
\put(-0.5,0){\includegraphics[width=4in]{#1}}
\put(2.4,0){\includegraphics[width=4in]{#3}}
\put(0.56,1.72){\includegraphics[width=23pt]{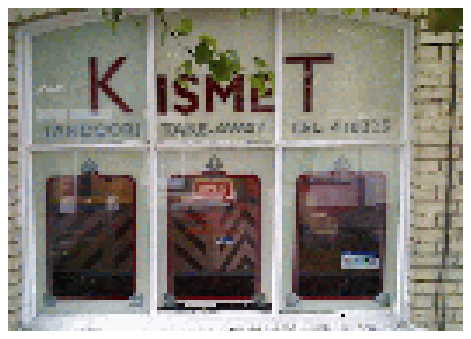}}
\put(3.46,1.72){\includegraphics[width=23pt]{kismet}}
\put(-0.05,2){(a)}
\put(0.125,0.44){\includegraphics[width=2.583in]{#2}}
\put(2.8,2){(b)}
\put(3.023,0.44){\includegraphics[width=2.583in]{#4}}
\end{picture} 
}

\newcommand{\fourgraphs}[4]{%
\unitlength=1in
\begin{picture}(6,4)(0,0)
\put(-0.5,2){\includegraphics[width=3.5in]{#1}}
\put(-0.1,4){(a)}
\put(2.8,2){\includegraphics[width=3.5in]{#2}}
\put(3.15,4){(b)}
\put(-0.5,-0.3){\includegraphics[width=3.5in]{#3}}
\put(-0.1,2){(c)}
\put(2.8,-0.3){\includegraphics[width=3.5in]{#4}}
\put(3.15,2){(d)}
\put(1.85,3.65){\includegraphics[width=20pt]{kismet}}
\put(1.85,1.45){\includegraphics[width=20pt]{kismet}}
\put(5.25,1.45){\includegraphics[width=20pt]{kismet}}
\put(5.25,3.65){\includegraphics[width=20pt]{kismet}}
\end{picture} 
}

\begin{document}
\maketitle

\section{Introduction}

Both ATLAS \cite{Collaboration:2011qk,daCosta:2011hh} and CMS
\cite{Khachatryan:2011tk} have now conducted searches for
supersymmetric particles with the 35 $fb^{-1}$ of $\sqrt{s}=7$ TeV
$pp$ collision data taken in 2010. All of these searches have involved
missing energy and jets, though the chosen techniques use different search variables and make differing requirements on (for example) the number of isolated leptons in each event.  None of
these searches has found a significant signal over the expected
Standard Model (SM) background, and so they have set limits on
sparticle production.  The most stringent limits come from the
ATLAS papers \cite{daCosta:2011hh,Collaboration:2011qk} each of which,
independently, subsumes the CMS exclusion \cite{Khachatryan:2011tk}
within its own.\footnote{Since both the ATLAS and CMS searches use
the same amount of data, it is reasonable to ask why the ATLAS reach
is so much greater than that of CMS\@.  The answer is that ATLAS and CMS
used very different experimental techniques.  The CMS search
\cite{Khachatryan:2011tk} was based on a {\em single}\/ cut on a 
variable called $\alpha_T$  
(along with cuts on the transverse hadronic momenta
of jets).  This variable is known to
strongly suppress QCD backgrounds, but is not designed with specific
kinematic properties in the supersymmetric (SUSY) signals in mind. The ATLAS
collaboration based its search on {\em four}\/ sets of cuts on {\em two}\/
different variables (the effective mass, $\meff$,
\cite{Hinchliffe:1996iu,Tovey:2000wk} and the stransverse mass
$\mttwo$, \cite{Lester:1999tx,Barr:2003rg,Cheng:2008hk}) which have
properties tailored more specifically to the kinematic properties of
$\squark\squark$, $\squark\gluino$ and $\gluino\gluino$ production.}
Within the CMSSM \citemsugraandcmssm{}, the strongest limit comes from the
ATLAS ``0-lepton'' search \cite{Collaboration:2011qk} which excludes equal mass squarks and
gluinos with masses below 775 GeV at 95\% C.L.~limits in the $A_0=0$,
$\tan(\beta)=3$, $\mu>0$ slice of CMSSM\@.  The equivalent limit from
the ATLAS ``1-lepton'' paper \cite{daCosta:2011hh} is 700 GeV, while
the CMS exclusion reaches 600 GeV.
Data taken in 2011 is expected to extend this reach, assuming lack of any
signal, up to squark and gluino masses of around 1000 GeV~\cite{Bechtle:2011dm}.

It is the aim of this paper to assess the impact of the ATLAS 0-lepton
results on the regions of the CMSSM favoured by indirect constraints and astrophysical data.  We note that an earlier study \cite{Allanach:2011ut} performed a similar update in response to the CMS results. In that study the likelihood function of the CMS exclusion was calculated by full Monte Carlo
simulation of LHC collisions.  We note also that in a later study,
\cite{Buchmueller:2011aa}, informed guesses for the forms of the likelihood
functions of the experiments (intended to be accurate in the vicinity of the CMS 95$\%$
exclusion contour) were used to 
update global frequentist fits to the CMSSM and other constrained
models, with similar conclusions. An
analysis of the ATLAS 1-lepton search result was also included in \cite{Buchmueller:2011aa}.

Here, we shall only consider the ATLAS 0-lepton result, since it is
more constraining than the other previous search results mentioned
above. A small amount of additional constraining power could, in
principle, be obtained by including the results of the other searches, but at the cost of
significant complication to our analysis.

Global fits allow a good fit in one observable to be traded for a somewhat
poor fit in a different observable in a statistically balanced way.
Fits to constrained SUSY models typically use 
the anomalous magnetic moment of the muon, the dark
matter relic density, direct searches for supersymmetric particles and Higgs
bosons, and electroweak observables to constrain simple SUSY models
simultaneously~\cite{Allanach:2005kz,Allanach:2006jc,Trotta:2006ew,Allanach:2006cc,Allanach:2007qk,Roszkowski:2007va,Allanach:2008iq,Feroz:2008wr,Buchmueller:2008qe,Trotta:2008bp,Roszkowski:2009sm,Feroz:2009dv,LopezFogliani:2009np,Roszkowski:2009ye,Buchmueller:2009fn,Buchmueller:2009ng,Baer:2010ny,Buchmueller:2010ai}.  
Variations with respect to all of the parameters
of the model that have an
impact on the predictions of the observables, including Standard Model (SM) parameters, are taken
into account. 
Various
algorithmic tools have now been developed 
to allow such a sampling of a multi-dimensional parameter space, which may be
multi-modal~\cite{Allanach:2007qj,Feroz:2008xx,Feroz:2011bj}. 
The lack of robustness of the results of such fits is illustrated by their large
prior
dependence~\cite{Allanach:2006jc,Allanach:2007qk,Trotta:2008bp,Allanach:2008iq}.
There are some predictions that are prior independent however, such as the
prediction of the lightest CP even Higgs mass, even in a fit to a 25 parameter
version of the minimal supersymmetric standard model (MSSM)~\cite{AbdusSalam:2009qd}. This is not surprising, since
LEP data provide a strong lower bound on the Higgs mass, and the model itself
imposes a close and strict upper bound. Thus, the data are constraining enough
themselves to dominate the prediction. 
Frequentist fits to edge measurements from hypothetical LHC SUSY edge
measurements showed an
incorrect confidence level (C.L.)
coverage of frequentist fits when the C.L.s are calculated by assuming a
$\chi^2$ distribution~\cite{Bridges:2010de}. There are no published coverage
studies of global SUSY fits and, since they are expected to be less robust
than fits to an LHC SUSY signal, a coverage study of the frequentist fits is
necessary and long overdue. A fit to a large volume string model with only 
two free parameters additional to the SM (the ratio of the Higgs
vacuum expectation values, $\tan \beta$, and an overall supersymmetry breaking
mass scale) did display approximate prior independence~\cite{Allanach:2008tu}.  
On the other hand, a fit to a model with three parameters additional to the SM
(minimal anomaly
mediated supersymmetry breaking) showed significant prior
dependence~\cite{AbdusSalam:2009tr}.  
Fits to models with more
than three additional parameters
have also (so
far) shown a lack of robustness~\cite{Roszkowski:2009sm,LopezFogliani:2009np,AbdusSalam:2009tr,AbdusSalam:2009qd}. 

Despite the lack of robustness of the global fits, we still find it
interesting to examine the effect of the recent search on them. Much of this
effect (ruling out light sparticles) is a robust property of the data rather
than of the model, and as such will be prior independent. It is useful to see
to what extent the experiments are able to rule out good-fit portions of the
models. Here, we exemplify in the CMSSM, a well-studied and well
defined model that has phenomenological properties that many other
supersymmetry breaking patterns will
follow. Specialising to the CMSSM allows us to take advantage of published
ATLAS 0-lepton signal rates in Ref.~\cite{auxmaterialplots}, which include next-to-leading order corrections
and detector effects that we could only crudely approximate were we to
simulate the events ourselves. 

The paper proceeds as follows: in Section~\ref{sec:0lep}, we review the basic
properties of the ATLAS 0-lepton search. In Section~\ref{sec:lm}, we describe
how we take into account correlated systematic background and signal errors in
order to provide a marginalised likelihood for the ATLAS 0-lepton search. The
likelihood is then validated against the 95$\%$ C.L.\ exclusion contours
published by ATLAS\@. We present the effect of the ATLAS 0-lepton search on global
fits in Section~\ref{sec:fits}, finishing with a summary and conclusions in
Section~\ref{sec:summ}. 

\section{The ATLAS 0-lepton Search\label{sec:0lep}}

\input{atlas-cut-table}

The cuts defining the search regions used by the ATLAS 0-lepton analysis are
given in Tab.~\ref{tab:atlascuts}. Also shown for each signal region $i \in
\{$A,B,C,D$\}$ are the number of observed events $n_o^{(i)}$ that made it past
cuts 
and the expected Standard Model backgrounds $n_b^{(i)}$ together
with their systematic errors $\sigma_b^{(i)}$. The $\sigma_b^{(i)}$ are
calculated by adding the uncorrelated background systematic and the jet energy
scale systematic in quadrature.
At each point in their model
grids, ATLAS also detailed the predicted number of signal events $n_s^{(i)}$
in each signal region. 

ATLAS constructed frequentist exclusion regions in SUSY parameter space using
a profile  
likelihood ratio method, taking into account theoretical and detector systematics and using
Monte Carlo toys to compute the coverage on a pair of SUSY model grids. The information from the four signal regions was
combined by defining the test statistic of each parameter point to be a
likelihood ratio 
given by the signal region demonstrating the best expected sensitivity to new
physics. 
Results were presented as 95\% confidence exclusion regions in the
$(m_{\tilde{g}},m_{\tilde{q}})$ plane for $m_{\chi_1^0}=0$ and in the $\tan
\beta=3$, $A_0=0$, $\mu>0$ slice of the CMSSM~\cite{Collaboration:2011qk}. 
95\% confidence exclusion regions (and expected sensitivity curves) were
also produced in Ref.~\cite{auxmaterialplots} for each signal region
individually. We shall use these curves to validate our statistical
calculation of the ATLAS 0-lepton search likelihood. 

\section{ATLAS 0-lepton Search Likelihood Map \label{sec:lm}}

ATLAS provides signal numbers throughout the CMSSM $m_0-m_{1/2}$ (the
GUT scale universal scalar and universal gaugino mass, respectively) plane for
the ratio of Higgs vacuum expectation values 
$\tan \beta=3$ and SUSY breaking scalar trilinear coupling $A_0=0$, obviating
the need for 
us to perform a SUSY signal event  
simulation in order to calculate the ATLAS 0-lepton search likelihood of each
point in the slice of parameter space. 
Given the information $\vec{\Sigma}^{(i)} =
(n_s^{(i)},n_b^{(i)},\sigma_s^{(i)},\sigma_b^{(i)})$  for a particular CMSSM
point and signal region $i$, we can model the expectation value for the number
of events observed in data as  
\begin{equation}
\lambda(\vec\Sigma^{(i)},\delta_s,\delta_b) 
	= n_s^{(i)} (1 + \delta_s \cdot \sigma_s^{(i)}) + n_b^{(i)} (1 + \delta_b \cdot \sigma_b^{(i)}),
\end{equation}
where the impact of systematic variations is accounted for by the nuisance
parameters $\delta_s, \delta_b$. We have neglected the luminosity error, which
is subdominant compared to the errors we include. 
The probability of observing $n$ events from a Poisson process which is expected to generate, on average, a mean of $\mu$ events, is given by
\begin{equation}
\mathrm{Poiss} \left( n|\mu \right)= \frac{e^{-\mu} (\mu)^n}{n!}. \label{likelihood}
\end{equation}
Taking the nuisance parameters to have Gaussian probability distribution
functions, the probability of 
observing $n_0^{(i)}$ events, with systematic deviations $\delta_s,\delta_b$ from
the central value is given by 
\begin{equation}
{\mathrm P}_{\text{syst}} ( n_o^{(i)}, \delta_s, \delta_b | \vec\Sigma^{(i)})
	=  \frac{1}{N^{(i)}} \mathrm{Poiss} \left( n_o^{(i)} | \lambda(\vec\Sigma^{(i)},\delta_s,\delta_b) \right)
			e^{-\frac{1}{2} (\delta_b^2 + \delta_s^2)},
\label{eqn:Psyst}
\end{equation}
where we have truncated the Gaussian modelling of the systematic errors at 
5$\sigma$ for convenience  (restricted to keep the signal and background contributions independently non-negative), leading to 
the normalisation factor 
\begin{equation}
N^{(i)} = \int_{\max(-5, -1/\sigma_s^{(i)})}^{5} \ d \delta_s
	\int_{\max(-5, -1/\sigma_b^{(i)})}^{5} \ d \delta_b \ 
	e^{-\frac{1}{2} (\delta_b^2 + \delta_s^2)}.
\label{eqn:norm}
\end{equation}
We then calculate the probability of observing $n_o^{(i)}$ events
\begin{equation}
  {\mathrm P}_m (n_o^{(i)} | {\vec\Sigma}^{(i)} ) =\int_{{\max(-5,-1/\sigma_s^{(i)})}}^{5} d \delta_s 
  	\int_{{\max(-5, -1/\sigma_b^{(i)})}}^{5} d \delta_b\ {\mathrm P}_{\text{syst}} 
		(n_o^{(i)},\delta_s,\delta_b).
\end{equation}

To validate the likelihood model and signal systematic estimation, we first
compute exclusion limits corresponding to the ATLAS expected and observed
results in the individual signal regions C and D. The inclusive di-jet signal
regions A and B are neglected for the purposes of this paper, as their
contributions to the constraints on the CMSSM parameter space were
sub-dominant.

At each model point considered by ATLAS for a single signal region $i$, we
compute 
the exclusion $p$-value defined as the cumulative marginalised likelihood for
$n_o^{(i)}$ observed events 
\begin{equation}
p_\text{excl}(n_0^{(i)}) = \sum_{n=0}^{n_o^{(i)}} {\mathrm P}_m (n | {\vec\Sigma}^{(i)}).
\end{equation}
This corresponds to the likelihood that the observed event count was given by
a downwards fluctuation from the Poisson mean of the nominal signal
hypothesis. The 95\% C.L. contour corresponding to 
$p_\text{excl}=0.05$
is then interpolated in the $m_{0}-m_{1/2}$
plane. 

Having determined suitable estimates of the signal uncertainties, we compute a
combined likelihood function that incorporates the measurements from the two
signal regions C and D. Here we diverge from the ATLAS strategy of taking the
single optimal signal region to determine the likelihood at each model point. 

We are using as our ``data'' the number of events $n_o^{(C)}$ passing cuts in ATLAS
signal region C, and the number of events $n_o^{(D)}$ passing cuts in ATLAS signal
region D.  To emphasise this, we notate our data as $\vec n$ where $\vec n
=(n_o^{(C)},n_o^{(D)})$.  The events in region D are a subset of those in region C (since
the only difference between the regions is that one has a harder cut on the
effective mass) so we note that the independent data that we are working with
are actually the numbers $n_o^{(D)}$ and $n_o^{(C)}-n_o^{(D)}$.
If the expected numbers of events passing the cuts 
in regions C and D are denoted by $\lambda_C$ and $\lambda_D$ respectively, then the
probability of observing our data $\vec n$ as a function of $\vec \lambda =
(\lambda_C,\lambda_D)$ is thus given by: 
\begin{equation}
P(\vec n | \vec \lambda) = \mathrm{Poiss}(n_o^{(D)}|\lambda_D)\ 
	\mathrm{Poiss}(n_o^{(D)}-n_o^{(C)}|\lambda_C-\lambda_D).
\label{eqn:combPoisson}
\end{equation}

Again, we can model the systematic uncertainties in the Poisson means,
\begin{align}
\lambda_C &= \lambda( {\vec\Sigma}^{(C)}, \delta_s, \delta_b), \\
\lambda_D &= \lambda( {\vec\Sigma}^{(D)}, \delta_s, \delta_b), 
\end{align}
where we keep the same $\delta_s$ and $\delta_b$ in both definitions, as we
assume that the uncertainties 
are fully correlated between the two signal regions. The probability of measuring
data $\vec n$  
incorporating systematic variations is then, using Eq.~\ref{eqn:combPoisson} and by analogy with Eq.~\ref{eqn:Psyst},
\begin{eqnarray}
{\mathrm P}_{\text{syst}} 
( \vec n, \delta_s, \delta_b | {\vec\Sigma}^{(C)}, {\vec\Sigma}^{(D)})
&=&  \frac{1}{N^{(C,D)}} \mathrm{Poiss} \left( n_o^{(C)} |
\lambda({\vec\Sigma}_C,\delta_s,\delta_b) \right) \times \\
		&& \mathrm{Poiss} \left( n_o^{(C)} - n_o^{(D)} | \lambda({\vec\Sigma}_C,\delta_s,\delta_b) - \lambda({\vec\Sigma}_D,\delta_s,\delta_b) \right)
			e^{-\frac{1}{2} (\delta_b^2 + \delta_s^2)} \nonumber
\end{eqnarray}
with the normalisation factor $N^{(C,D)}$ defined similarly to \ref{eqn:norm} as
\begin{equation}
N^{(C,D)} = \int_{\max (-5, -1/\max (\sigma_s^{(C)}, \sigma_s^{(D)}))}^{5} d \delta_s\
 \int_{\max (-5, -1/\max ( \sigma_b^{(C)}, \sigma_b^{(D)}))}^{5} d \delta_b\
	e^{-\frac{1}{2} (\delta_b^2 + \delta_s^2)}.
\end{equation}
Marginalising over the systematics once more produces the probability
of measuring  $\vec n$ under the nominal signal hypothesis,
\begin{eqnarray}
{\mathrm P}_m (\vec n |  {\vec\Sigma}^{(C)}, {\vec\Sigma}^{(D)}) &=&
\int_{\max (-5, -1/\max (\sigma_s^{(C)}, \sigma_s^{(D)}))}^{5} d \delta_s\
 \int_{\max (-5, -1/\max ( \sigma_b^{(C)}, \sigma_b^{(D)}))}^{5} d
 \delta_b\ \{ \nonumber \\
&&   {\mathrm P}_{\text{syst}} (\vec n,\delta_s,\delta_b |
   {\vec\Sigma}^{(C)}, {\vec\Sigma}^{(D)}) \}. \label{combl}
\end{eqnarray}
We shall refer to ${\mathrm P}_m(\vec n | {\vec\Sigma}^{(C)}, {\vec\Sigma}^{(D)})$ as the likelihood, or the ATLAS
0-lepton search likelihood, in what follows. 

\subsection{Validation of the search likelihood penalty}
\label{sec:validation}

\begin{figure}\begin{center}{\unitlength=1in
\begin{picture}(6,3)(0,0)
\put(0,0){\includegraphics[width=3in]{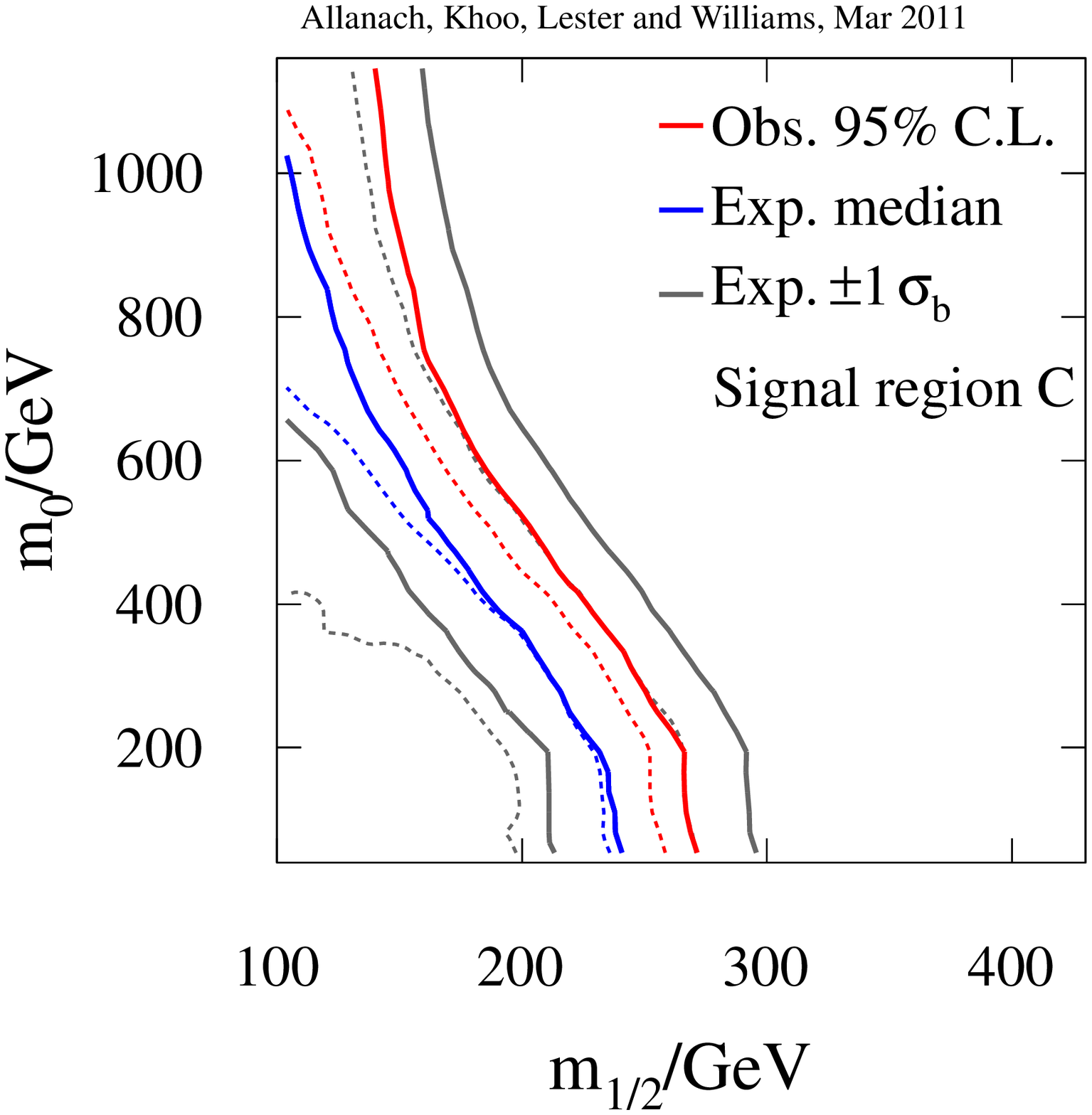}}
\put(0,2.8){(a)}
\put(3,0){\includegraphics[width=3in]{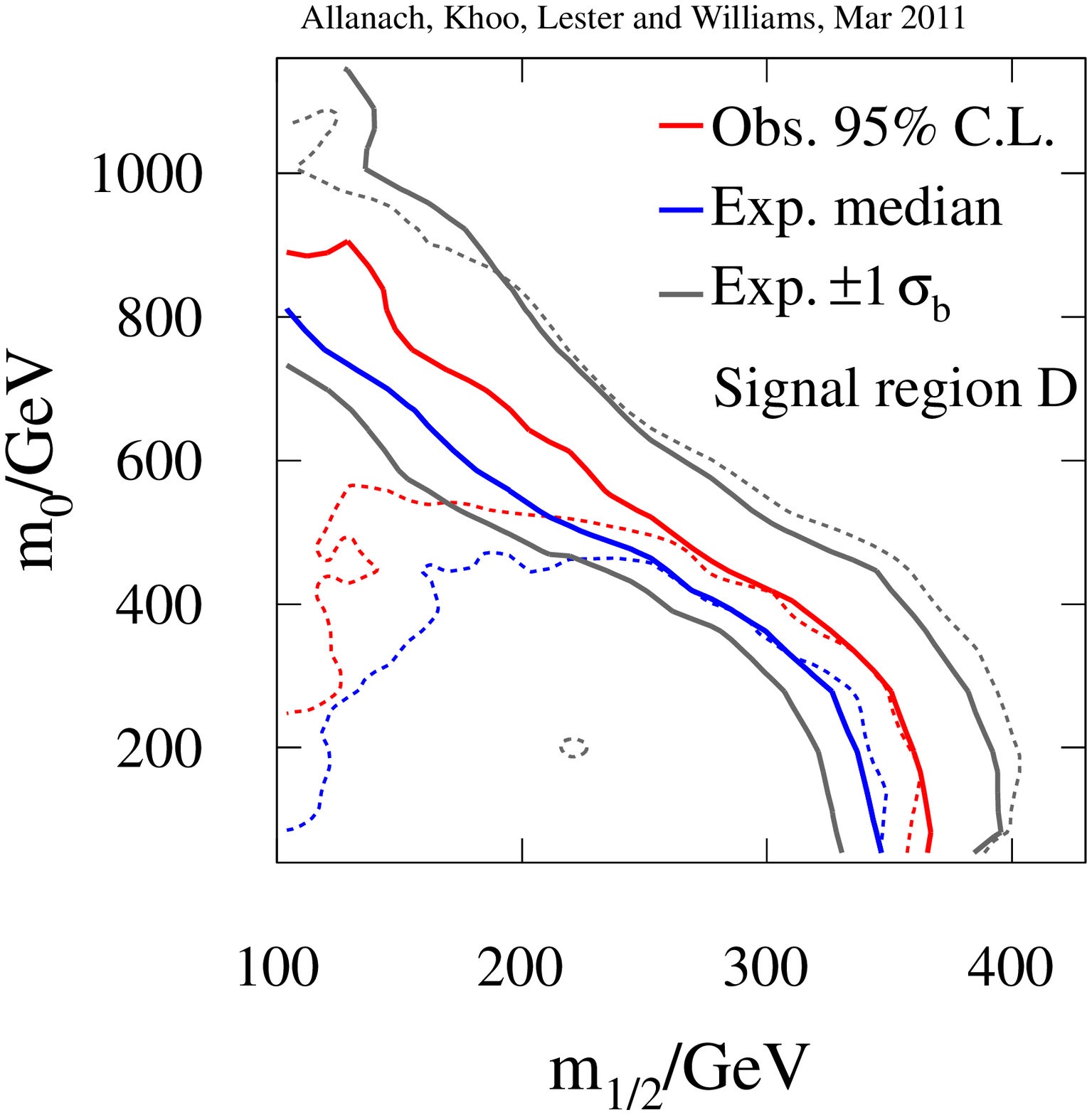}}
\put(3,2.8){(b)}
\put(2.2,1.5){\includegraphics[width=30pt]{kismet}}
\put(5.2,1.5){\includegraphics[width=30pt]{kismet}}
\end{picture} 
\caption{Validation of our statistical analysis of the ATLAS
  0-lepton search likelihood.  
We reproduce the ATLAS expected and observed 95\% C.L. 
  limits from the 0-lepton search for (a) signal region C and (b) signal
  region D. Solid (dashed) lines    indicate our (ATLAS') exclusion contours. 
  \label{fig:valid}}  
}\end{center}\end{figure}

We validate our statistical framework (defined in the previous section) by
attempting to reproduce the official ATLAS exclusion limits from
Ref.~\cite{Collaboration:2011qk}. 
A systematic error\footnote{$\sigma_s^{(i)}$ accounts
for higher order corrections and jet energy scale uncertainties among others.}
$\sigma_s^{(i)}$ on the signal was used in the ATLAS 
results, but the values of $\sigma_s^{(i)}$ were not made public. 
In order to account for signal systematics, 
we vary $\sigma_s^{(C)}$ and $\sigma_s^{(D)}$ in order to provide a reasonable
fit to  
the official ATLAS 95$\%$ C.L.\ exclusion contours in the parameter regions most
sensitive to the global fit. Varying them manually, we find that
$\sigma_s^{(C)}=0.6$ and
and $\sigma_s^{(D)}=0.3$ respectively, provide a reasonable fit in the most
important area of the parameter plane for each signal region. 
We find that the exclusion
contours are not so sensitive to the 
precise values of $\sigma^{(C)}$ and $\sigma^{(D)}$: changing either by 0.05 
moves the contours almost imperceptibly.
The exclusion contours are 
interpolated in the $m_0-m_{1/2}$ plane after computing $p_\text{excl}(n_o)$
at each model 
point for signal regions C and D separately (see Fig.~\ref{fig:valid}). 

Our solid exclusion contours are seen to match the official ATLAS
dashed contours well, particularly at low $m_0$ and high $m_{1/2}$, i.e.\ in the
lower right-hand corner of the plots. This is crucial, since the global fits
before including
ATLAS 0-lepton search results favour this region of the CMSSM parameter space,
as is demonstrated below in Fig.~\ref{fig:m0m12}, and hence the ATLAS search
likelihoods will have the greatest impact in this region. 
Our approximations do very well at high $m_{1/2}$, close to the favoured CMSSM point. 
Elsewhere, at lower values of $m_{1/2}$ and high $m_0$, the approximation is
less good, particularly in signal region D. This is likely due to the
assumption of a signal uncertainty that, within each signal region, is flat. 
In fact, the poor signal region D likelihood reproduction in the small
$m_{1/2}$ larger $m_0$ area will not make much difference to our combined
likelihood, since there it is dominated by signal region C anyway, where our
approximation is reasonable, as Fig.~\ref{fig:valid}a shows. 

\begin{figure}\begin{center}{\unitlength=1.5in
\begin{picture}(3,2)(-0.2,0)
\put(-0.5,0){\includegraphics[width=5.25in]{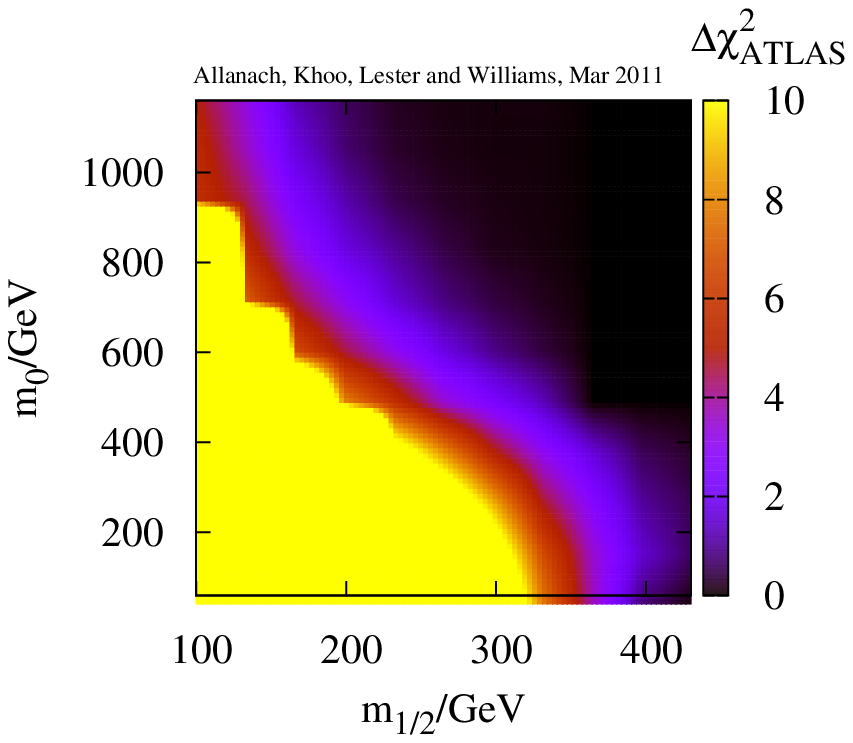}}
\put(0.125,0.44){\includegraphics[width=3.39in]{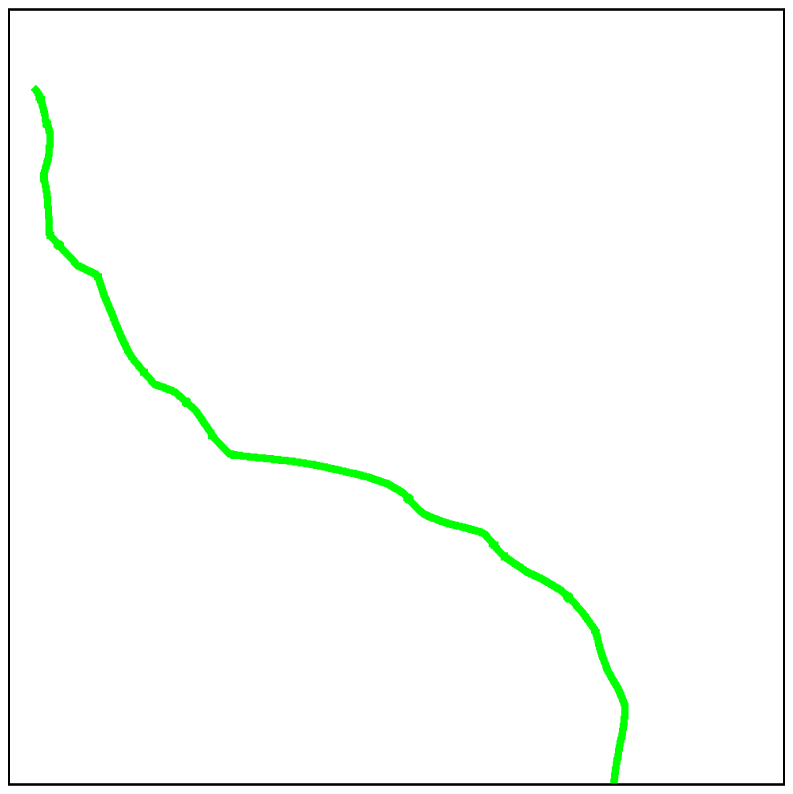}}
\put(1.67,1.72){\includegraphics[width=30pt]{kismet}}
\end{picture} 
\caption{Our approximation to the ATLAS 0-lepton search CMSSM likelihood map
  for $\tan \beta=3$, $A_0=0$. We display 
  $\Delta \chi^2_{ATLAS}$ as the
  background colour density. 
  The ATLAS 95$\%$ C.L. exclusion limit is shown as the light (green)
  solid line. 
  \label{fig:scan}}  
}\end{center}\end{figure}

For the jets plus $\mpt$ search (0-lepton), where the signal involves just
high energy jets and  missing transverse momentum, 
we expect the signal rate to be approximately independent
of $\tan \beta$ and $A_0$. This is because the signal is dominated by the
strong interaction cross-sections of squark and gluino production, which do
not depend 
to any significant degree on those parameters. 
The accuracy of the $\tan \beta-A_0$ independence assumption was explicitly
checked recently in the CMS $\alpha_T$ search~\cite{Allanach:2011ut}, which is
also looking for events with jets and missing transverse momentum. It was found that, for CMSSM 
global fits,  the CMS $\alpha_T$ search likelihood is well approximated by
ignoring any $A_0$ or $\tan \beta$ dependence. 
Being able to model the dependence of the ATLAS search likelihood on
$m_0$ and $m_{1/2}$, 
while ignoring the effect of $A_0$ and $\tan \beta$ leads to
a significant simplification when we come to take it into account in our
global CMSSM fits.  We shall neglect $A_0$ and $\tan \beta$ dependence.

We investigate the combination of signal regions C and D into P$_m$ in
Fig.~\ref{fig:scan}. 
 \begin{equation}
\Delta \chi^2_{ATLAS}=-2 \ln (\mbox{P}_m/ \mbox{P}_m(0\mbox{~sig}))
\end{equation}
is shown as the   background colour density, where P$_m(0\mbox{~sig})$ is the combined
  0-lepton search likelihood penalty from Eq.~\ref{combl} in the limit of no
  signal events. 
We see that the shape of the background colour density closely follows the
shape of the official ATLAS exclusion limit. It is also around $\Delta
\chi^2=5.99$, which would be the 95$\%$ C.L.\ exclusion in the limit of 
Gaussian statistics\footnote{We note here that in some regions of parameter
  space, the event numbers are very small and so one cannot use the
  Gaussian limit.}. 
For increasing $m_{1/2}$ GeV, we see $\Delta \chi^2$ reaching a 
constant in Fig.~\ref{fig:scan} because there is no SUSY signal, since squarks
and gluinos become too
heavy to be produced. 
At large $m_0$ and small $m_{1/2}$, the SUSY signal is strongly dominated by
gluino pair 
production, where the gluinos have three-body decays into squarks. Thus the
dependence of ${\mathrm P}_m$ on $m_0$, if it is above 1160 GeV, is negligible. 
We shall therefore model the likelihood as follows: we use  for
$m_{1/2}>430$ GeV, $n^C_s=n^D_s=0$. We also use this zero signal limit for 
$m_{1/2}>340$ GeV and $m_0>430$ GeV.
For $m_0>1160$ GeV and $m_{1/2}<430$
GeV, we use the 
${\mathrm P}_m$ value given by the $m_0=1160$ line on the figure. For
$m_0<1160$, $m_{1/2}<430$, we interpolate linearly within the grid of
$\Delta \chi^2_{ATLAS}$. 

\section{Global CMSSM Fits Including the ATLAS Search \label{sec:fits}}
We shall use a previous global Bayesian fit of the CMSSM from the {\tt KISMET}
(Killer Inference in Supersymmetric METeorology)
collaboration~\cite{Allanach:2007qk}
to: the relic density of dark matter, the anomalous magnetic moment of the
muon, previous direct searches for sparticles, the branching ratios $BR(b
\rightarrow s \gamma)$, $BR(B_s \rightarrow \mu\mu)$, $M_W$, $\sin^2
\theta_w^l$, as well as 95$\%$ exclusions from LEP and Tevatron direct search
data. The ranges of parameter considered were: $2 < \tan \beta < 62$,
$|A_0|/\mbox{TeV}<4$, $60<m_{1/2}/\mbox{GeV}<2000$, $60<m_0/\mbox{GeV}<4000$. 
Variations of the top mass, the strong coupling constant,
the fine structure constant and the bottom mass were all included. Various
different prior distributions were examined in Ref.~\cite{Allanach:2007qk},
but here we 
shall use the example of priors flat in the parameters listed above, except
for $m_0$ and $m_{1/2}$, which are flat in their logarithm. Using such log
priors allows us to illustrate the effects of the 0-lepton search more
acutely than with purely linear priors. Rigorous convergence criteria were
satisfied by the fits, which were performed by ten Metropolis Markov Chains
running simultaneously. For more details on the fits, we refer interested
readers to Ref.~\cite{Allanach:2007qk}.

We take 2.7 million points from the fits, whose densities in parameter space
are proportional to their posterior probability distributions. We then
re-weight each point by multiplying its global fit likelihood by ${\mathcal
  L}$ calculated from the 0-lepton search. By plotting the posterior
probability distributions before and after the re-weighting, we
then examine the effect of the ATLAS SUSY exclusion data on the CMSSM
fits. 

\begin{figure}\begin{center}\twographs{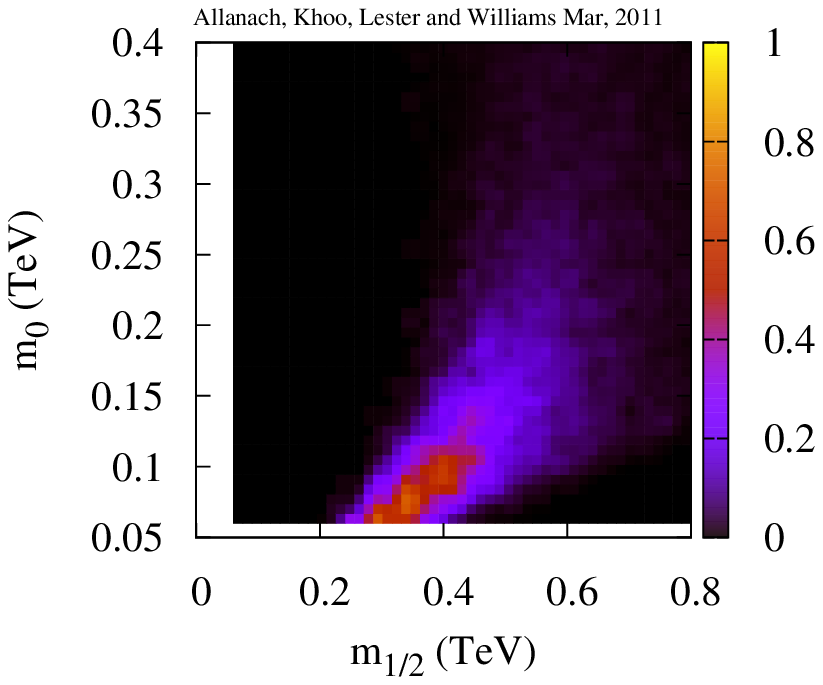}{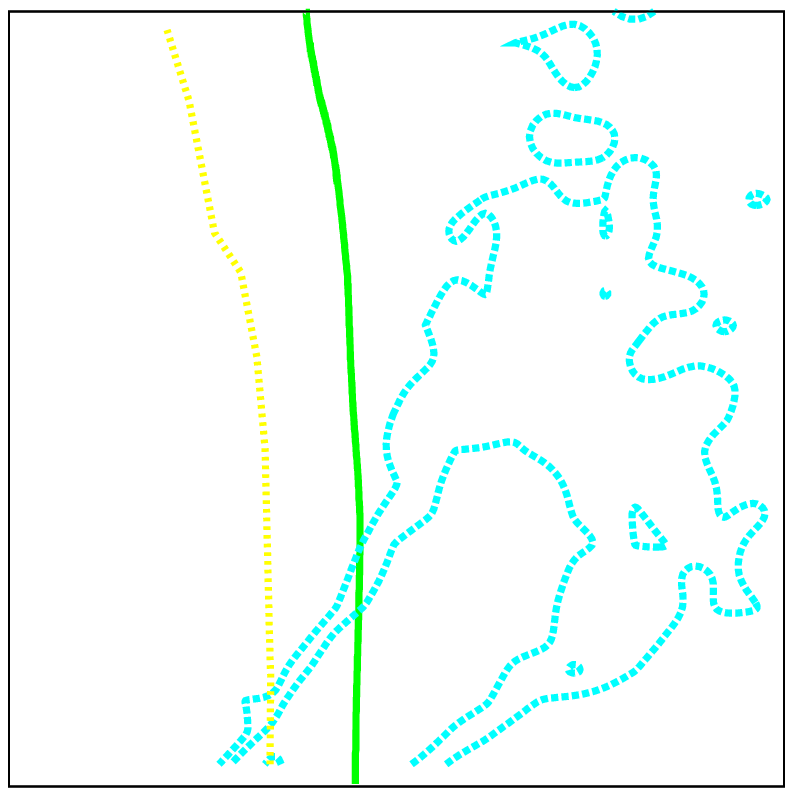}{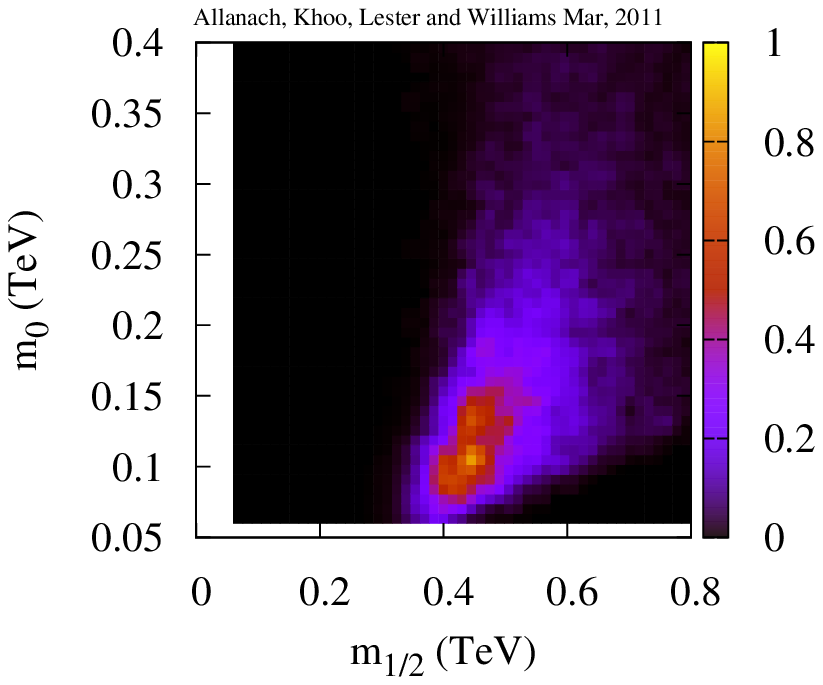}{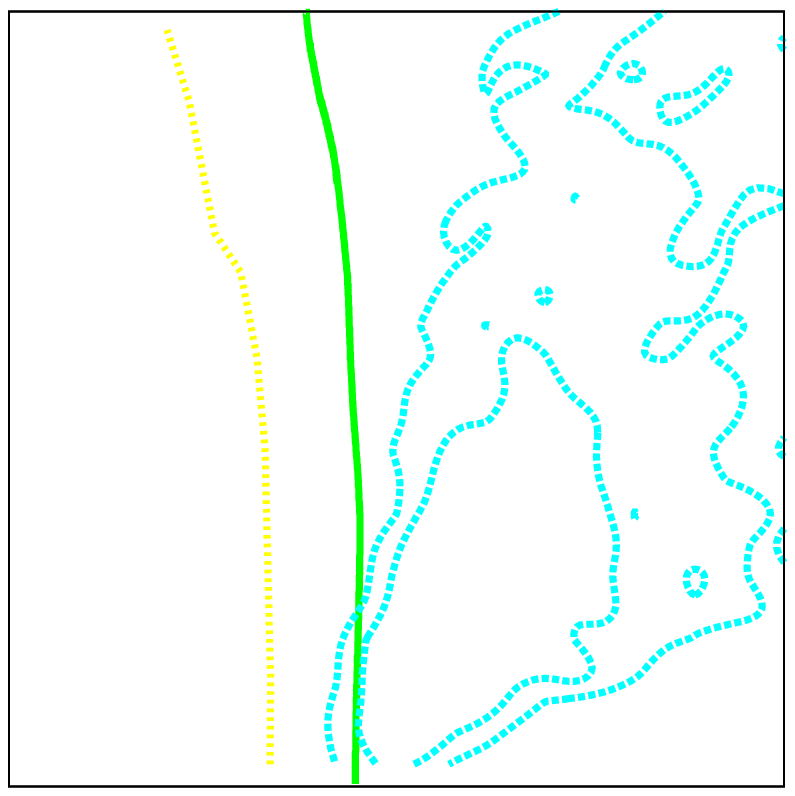}
\caption{Global CMSSM fits in the $m_0-m_{1/2}$ plane: (a) excluding the ATLAS 0-lepton search and 
  (b) including the ATLAS 0-lepton search likelihood. 
The posterior probability of each bin is shown as the background colour,
normalised to the maximum bin probability.
The region to the left of the almost vertical solid green (dotted yellow)
curve is excluded by the ATLAS 0-lepton search (CMS $\alpha_T$ search) at the
95$\%$ C.L. 
The cyan inner
(outer) contour 
shows the 68$\%$ (95 $\%$) Bayesian credibility region. \label{fig:m0m12}} 
\end{center}\end{figure}
We display the impact of the 0-lepton search on the $m_0-m_{1/2}$ plane in
Fig.~\ref{fig:m0m12}. Fig.~\ref{fig:m0m12}a shows that the search 95$\%$
contour covers much of the current region that fits indirect data well at low
values of $m_0$. We emphasise that we have used the full likelihood function 
and not just a simple cut based on the exclusion curve.
When our approximation to the likelihood function in
Fig.~\ref{fig:scan} is applied to the global fit, much of the probability mass moves
to higher values of $m_0$ and $m_{1/2}$, despite the fact
that the anomalous magnetic momentum of the muon would prefer somewhat lower
values. This effect is much more pronounced
in the ATLAS 0-lepton search than in the CMS $\alpha_T$ search
because of the more stringent exclusion of the ATLAS analysis, as shown by a
comparison between Fig.~\ref{fig:scan} and Fig.~3 of
Ref.~\cite{Allanach:2011ut}. 
The fits are performed under the
CMSSM hypothesis, so the total probability is conserved, even after the ATLAS
search data have been used to constrain the model. 
We could quantify the
difference the search has made to the Bayesian evidence of the model, but such
an inference is likely to not be robust until the CMSSM is strongly
constrained by supersymmetric signals. The non-robustness manifests as a high
degree of prior dependence in the evidence~\cite{AbdusSalam:2009tr}. 

\begin{figure}\begin{center}
\twographs{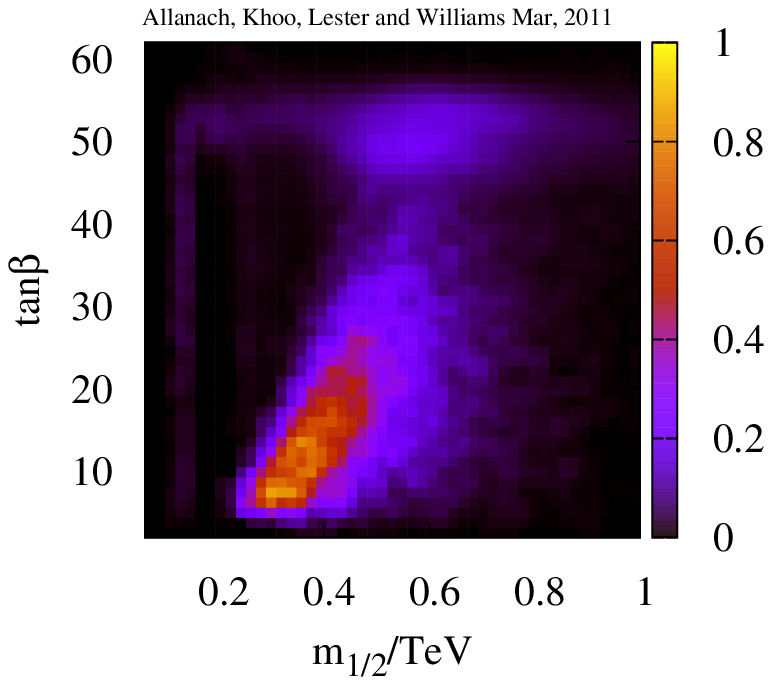}{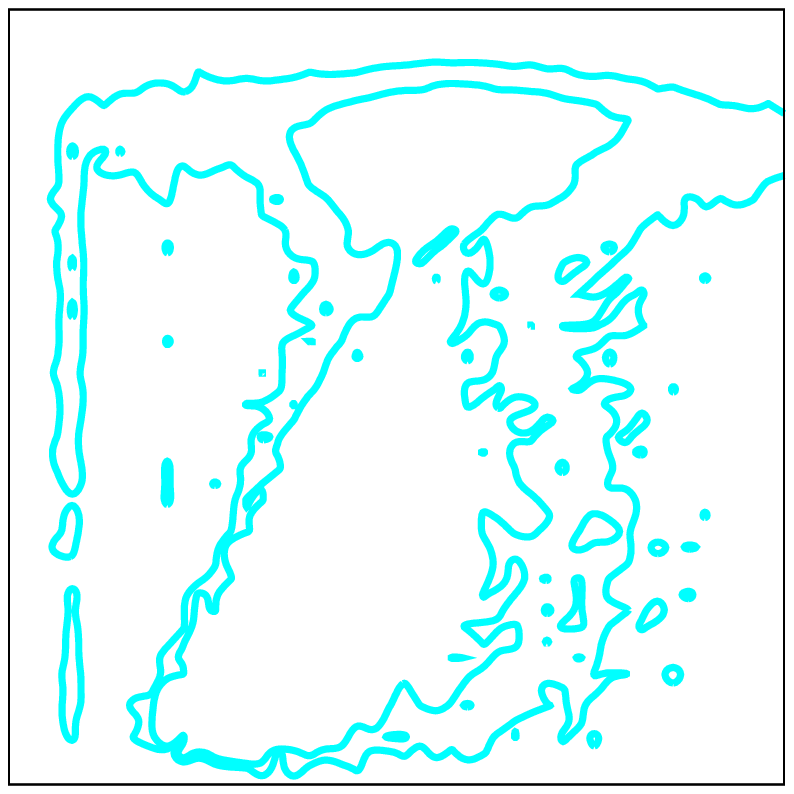}{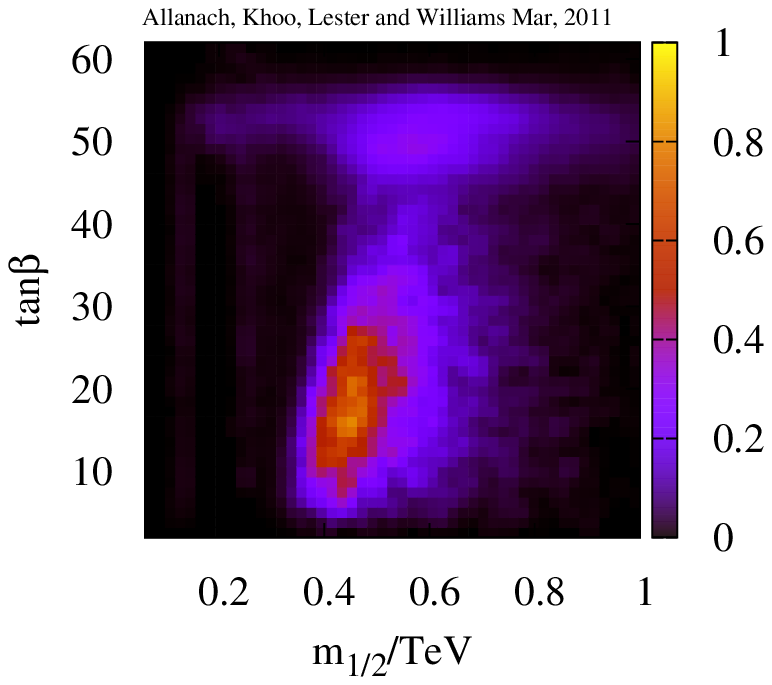}{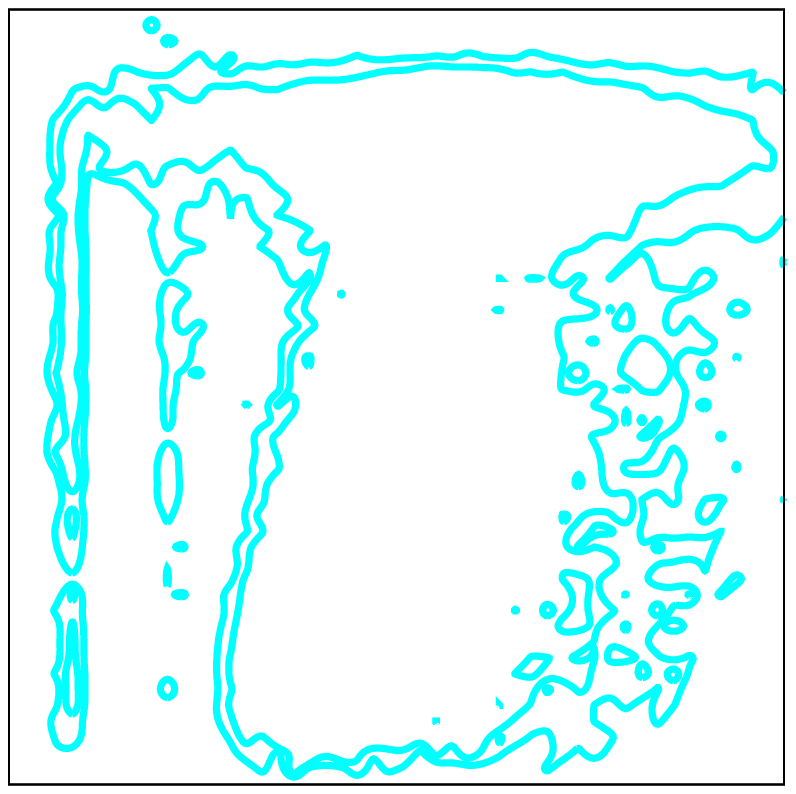}
\caption{Global CMSSM fits in the $m_{1/2}-\tan\beta$: (a) excluding the ATLAS
  0-lepton search and  
  (b) including the ATLAS 0-lepton search likelihood. 
The posterior probability of each bin is shown as the background colour,
normalised to the maximum bin probability.
The cyan inner
(outer) contour 
shows the 68$\%$ (95 $\%$) Bayesian credibility region. \label{fig:m12tb}}  
\end{center}\end{figure}
Fig.~\ref{fig:m12tb} displays the effect of the 0-lepton search on the
$m_{1/2}-\tan \beta$ plane. As well as moving the probability mass up in
$m_{1/2}$, we see that it is moved upward in $\tan \beta$ as well. This effect
is due to the positive correlation between $m_{1/2}$ and $\tan \beta$ evident in
Fig.~\ref{fig:m12tb}a in the high probability region. The correlation can be
understood as a consequence of the fits preferring a positive contribution to
the anomalous magnetic momentum of the muon, $\delta a_\mu$. 
$\delta a_\mu$ is proportional to $\tan
\beta/M_{SUSY}^2$, where $M_{SUSY}$ is the mass scale of sparticles in the loop
that contribute. Thus, if $M_{SUSY}$ is forced upward by the ATLAS search, to
get an equivalent $\delta a_\mu$, $\tan \beta$ must also increase. The
vertical arm at the left hand side of the plots corresponds to  the higgs pole
region, where neutralinos annihilate efficiently through an $s$ channel
lightest CP even higgs boson. The higgs pole region has low $m_{1/2}$ and
high $m_0$, and so isn't yet affected much by the ATLAS search.

\begin{figure}\begin{center}{\fourgraphs{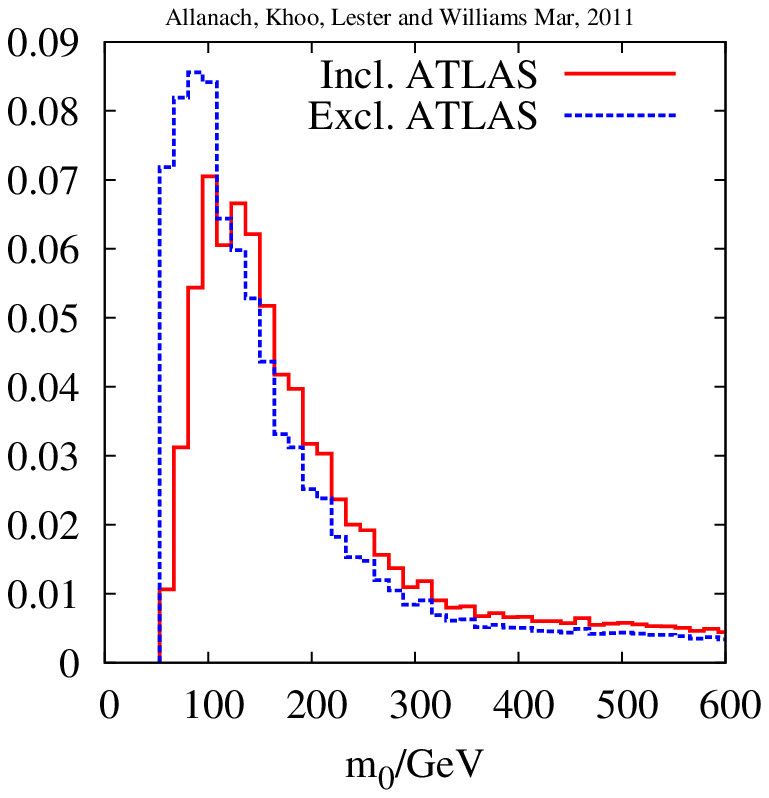}{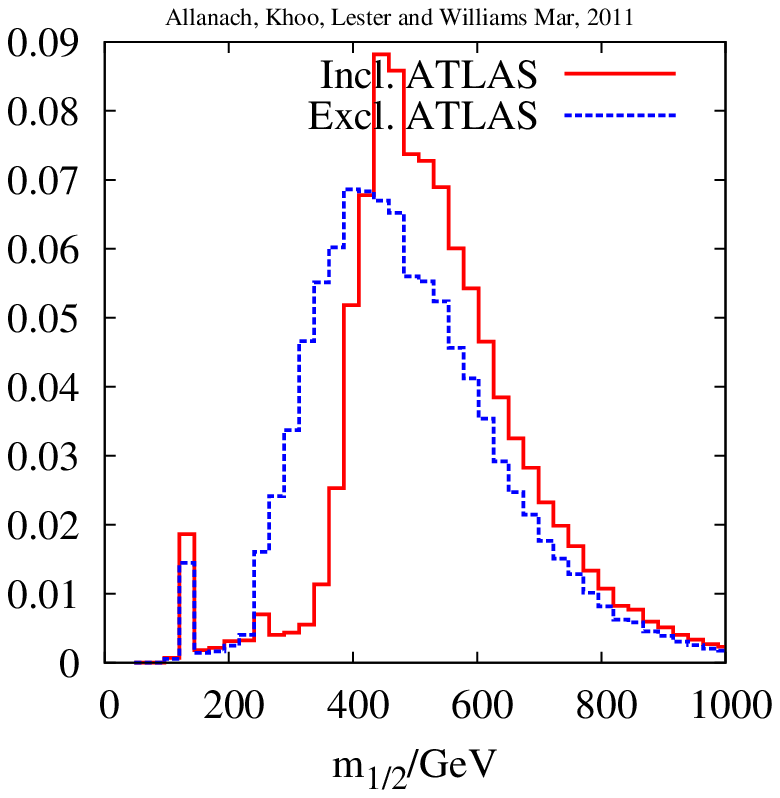}{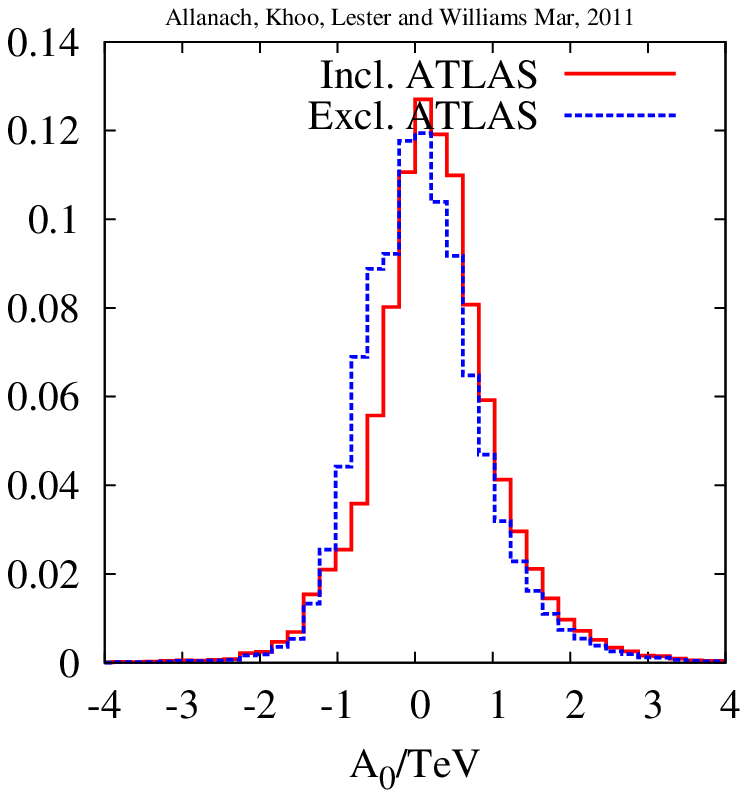}{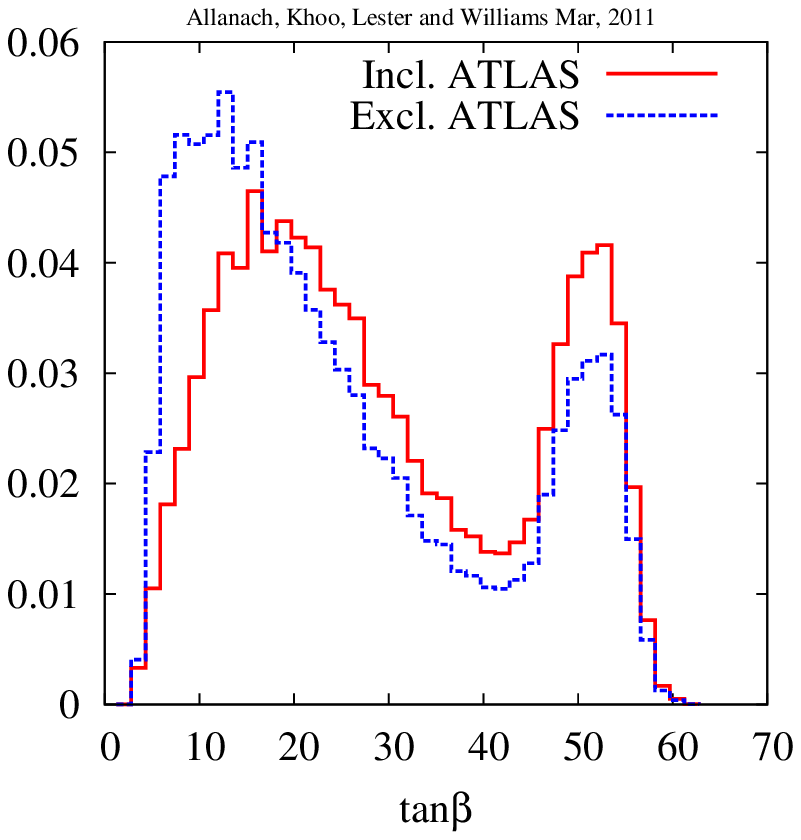}
\caption{Effect of the ATLAS 0-lepton, jets and missing momentum search \cite{Collaboration:2011qk} on one dimensional 
  probability distributions of CMSSM parameters. \noteOnAreasAndLabels
 \label{fig:params}}}\end{center}\end{figure}
The effect of the ATLAS 0-lepton search \cite{Collaboration:2011qk} on individual CMSSM parameters is
shown in Fig.~\ref{fig:params}, and can be understood in terms of the effect
of the search on the higher dimensional parameter space: $m_0$ and $m_{1/2}$
are pushed to larger values, as is $\tan \beta$ due to the correlations
mentioned above. There is almost no change in the probability distribution of $A_0$,
indicating that $A_0$ isn't strongly correlated in the global fits with the
other parameters.

\begin{figure}\begin{center}{\fourgraphs{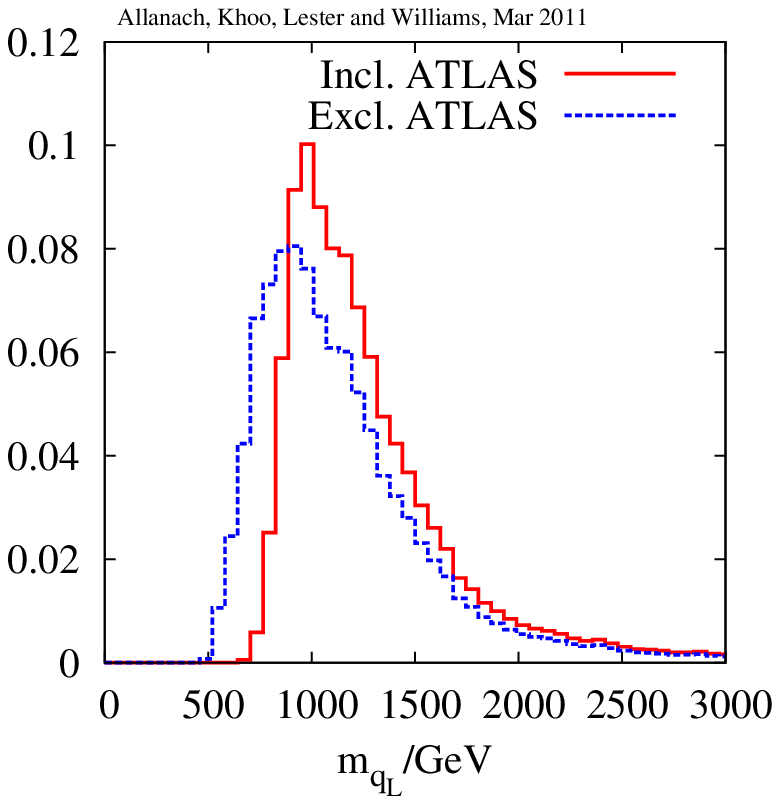}{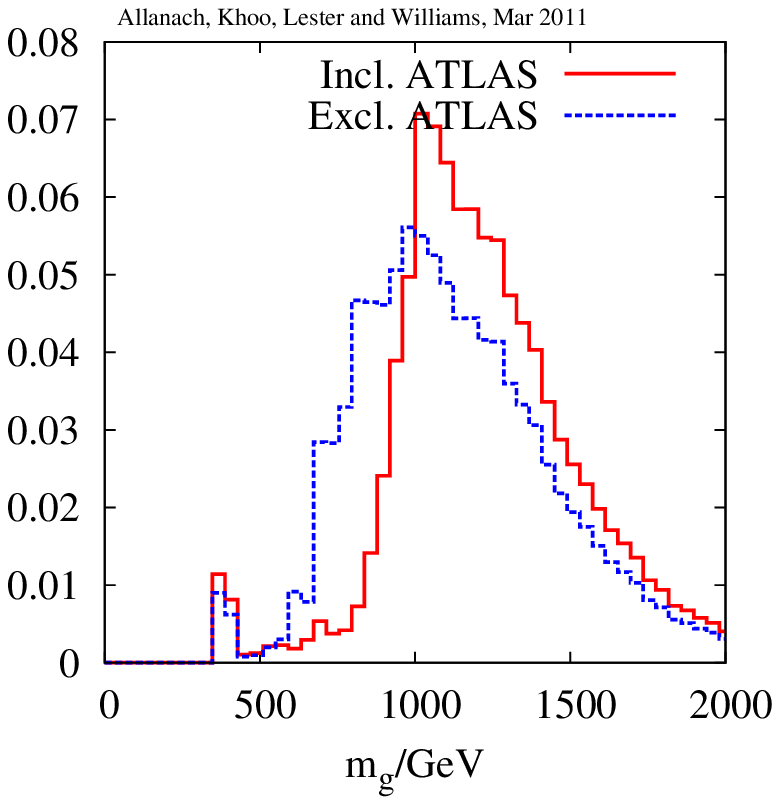}{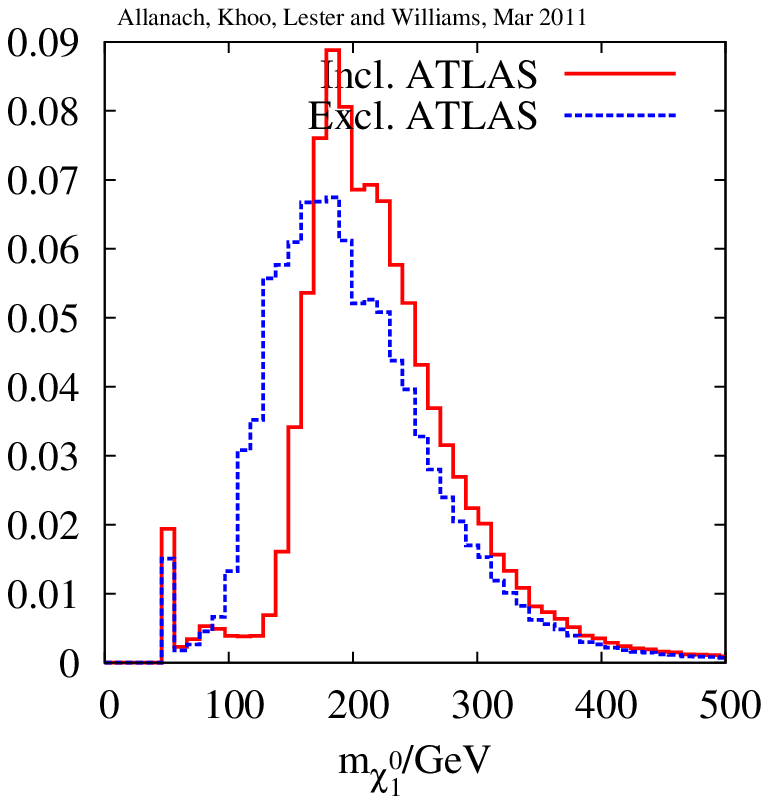}{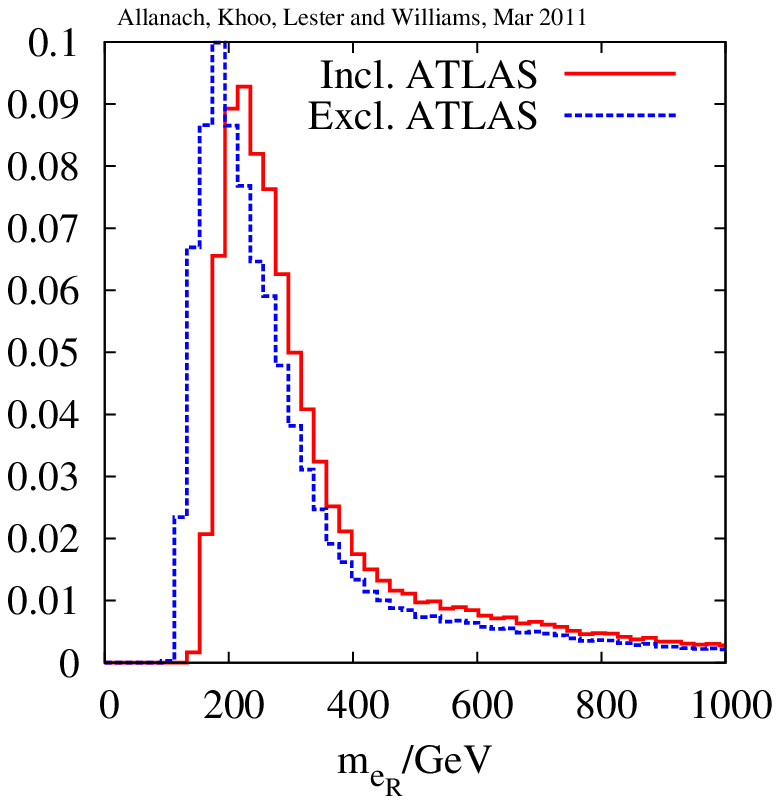}
\caption{Effect of the ATLAS 0-lepton, jets and missing momentum search \cite{Collaboration:2011qk} on the 
  probability distributions of sparticle masses in the CMSSM\@. \noteOnAreasAndLabels \label{fig:masses}}  
}\end{center}\end{figure}
A change in the probability distributions in the CMSSM parameters implies a change in the probability distributions of sparticle masses. We
display a representative sample of these in Fig.~\ref{fig:masses}. The squark
and gluino masses are predictably pushed to be heavier by the CMSSM search, as
is the neutralino, since in the CMSSM it is controlled by the same parameter
as the gluino mass ($m_{1/2}$), and is therefore highly correlated. We see a
similar effect for the right-handed slepton ${\tilde e}_R$, which is
strongly correlated with $m_0$ and is thus pushed to somewhat heavier values. 
Although in general, global fits are not expected to be robust until
significant SUSY signals are detected, the moving of sparticle masses to
heavier values by the ATLAS exclusion {\em is}\/ expected to be. 


\begin{figure}\begin{center}{\unitlength=1.1in
\begin{picture}(3,2)(0,0)
\put(0,0){\includegraphics[width=3.3in]{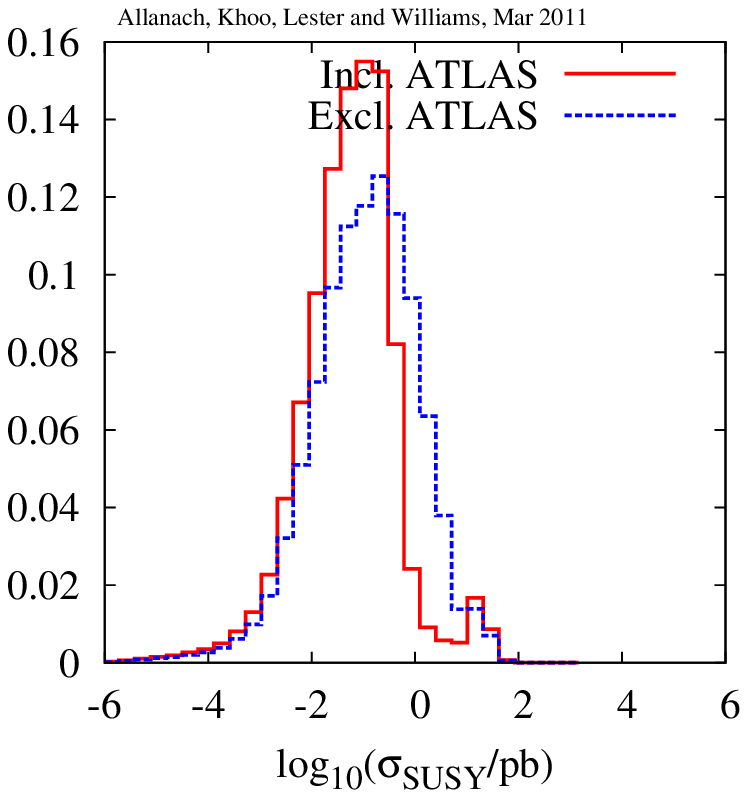}}
\put(2.02,1.45){\includegraphics[width=20pt]{kismet}}
\end{picture}
\caption{Effect of the ATLAS 0-lepton, jets and missing momentum search \cite{Collaboration:2011qk} on the total SUSY cross-section
  $\sigma_{SUSY}$ in the CMSSM in $pp$ collisions at $\sqrt{s}=7$ TeV. \noteOnAreasAndLabels  \label{fig:sigma}}
}\end{center}\end{figure}
We display the effect of the ATLAS 0-lepton search on the probability
distribution for the total production cross-section of sparticles
$\sigma_{SUSY}$ in Fig.~\ref{fig:sigma}. 
As expected, heavier sparticles
mean that the cross-sections decrease somewhat.

\section{Summary and Conclusions \label{sec:summ}}
Global fits of the CMSSM to indirect data provide us with a ``weather
forecast'' for future sparticle production, under the CMSSM hypothesis.
We use {\tt KISMET} fits that take into account
the anomalous magnetic moment of the muon, the dark
matter relic density  and electroweak observables and direct searches for
supersymmetric particles and Higgs bosons.
These data have the power to constrain approximately 
two free parameters additional to the SM: essentially, the dark matter relic
density constraint is  
strong enough to constrain one dimension, and the combination of all of the
other observables jointly constrains another. 
The ATLAS 0-lepton search \cite{Collaboration:2011qk} has significantly
extended previous exclusion limits in the CMSSM and it has a significant
effect on the global fits. 
The search bites off the part of the parameter space where both squarks and
gluinos are light, but also has 
some other non-trivial effects: for instance $\tan \beta$ is pushed to higher
values. A recent CMS $\alpha_T$ search~\cite{Khachatryan:2011tk} based on the
$\alpha_T$ variable,
picked because it was thought to be more robust with
respect to fluctuations in Standard Model backgrounds, also produced some of
these 
effects on the global fits~\cite{Allanach:2011ut}, although it had a slight
$\sim 1 \sigma$ 
excess in the number of events, meaning that intermediate sparticle masses
were somewhat preferred. Since the ATLAS 0-lepton search did not have such 
an excess, there is no relative preference for intermediate sparticle masses.
The ATLAS exclusion reaches further than the one produced by CMS, with a
consequently larger effect on the global fits. ATLAS had a different search
strategy, relying on more standard cuts on $m_{eff}$, $\mpt$ and $m_{T_2}$
which differ in different signal regions of parameter space and which have been
somewhat optimised to increase the constraining power of the search. 
The heavier sparticles implied by the ATLAS 0-lepton exclusion means that 
the weather forecast for LHC sparticle production is somewhat more arid: 
heavier sparticles have less phase space to be produced in the collisions, and
their production cross-section decreases. 
With a most likely
$\sigma_{SUSY}=0.1$ pb, 
there is still
plenty of 
opportunity for LHC sparticle production in the coming years.

ATLAS produced a useful amount of information about their search in
published auxiliary data, including backgrounds and uncertainties and 
expected signal rates throughout parameter space. 
The signal rates allowed
us to take their search into account without having to re-perform
event generation, which would be a CPU-time bottleneck and a significant
complication in the analysis. We are thus also able to perform the 
fits implicitly taking detector effects into account. 

\acknowledgments
This work has been partially supported by STFC\@. TJK is supported by a
Dr. Herchel Smith Fellowship from Williams College. We thank other members of
the Cambridge SUSY working group for discussions held.
\bibliographystyle{JHEP}
\bibliography{a}

\end{document}

%% file: ourSusydefs.tex
\newcommand{\meff}{\ensuremath{m_{\mathrm{eff}}}} 

\newcommand{\mttwo}{\ensuremath{m_\mathrm{T2}}}

%% file: atlas-cut-table.tex
\begin{table}
\begin{center}
\begin{tabular}{l|llll}
\hline
 & Region A & Region B& Region C& Region D \\
\hline
Number of required jets & $\geq 2$ & $\geq 2$  &$ \geq3 $  & $\geq 3$ \\
 Leading jet $p_{T}$ & $>120$ GeV & $>120$ GeV & $>120$ GeV & $>120$ GeV\\
 Subsequent jet(s) $p_{T}$& $>\phantom{8}40$ GeV & $>\phantom{8}40$ GeV& $>40$ GeV& $>\phantom{8}40$ GeV\\
 $E_{T}^{miss}$ & $>100$ GeV& $>100$ GeV& $>100$ GeV& $>100$ GeV\\
 $\Delta\phi$(jet,$\vec{P}_{T}^{miss})_{min}$ & $>0.4$& $>0.4$& $>0.4$& $>0.4$ \\
 $E_{T}^{miss}/m_{eff}$ & $>0.3$ & - & $>0.25$& $>0.25$ \\
 $m_{eff}$ & $>500$ GeV & - & $>500$ GeV& $>1000$ GeV\\
$m_{T2}$ & - & $>300$ GeV& - & - \\
\hline
Observed & 87 & 11& 66& 2\\
Standard Model background & 118$\pm$25$\pm$32 & 10$\pm$4.3$\pm$4 & 88$\pm$18$\pm$26 & 2.5$\pm$1$\pm$1 \\
\hline
\end{tabular}
\end{center}
\caption{The cuts used to define the four signal regions of the ATLAS-0-lep
  analysis \cite{Collaboration:2011qk}.  A veto on events containing
  isolated leptons with $p_T>20$ GeV is an additional requirement of
  Ref.~\cite{Collaboration:2011qk} but is not shown in the table. We also
  display 
  the number of events ATLAS observed in each region, along with the expected
  Standard Model backgrounds. The first uncertainty represents the
  uncorrelated systematic on the background, whereas the second labels the
  jet energy scale systematic.\label{tab:atlascuts}}
\end{table}